\documentclass[12pt]{iopart}

\usepackage{iopams}
\newtheorem{lemma}{Lemma}
\newtheorem{proposition}{Proposition}
\newtheorem{theorem}{Theorem}
\newtheorem{corollary}{Corollary}

\newcommand{\be}{\begin{equation}}
\newcommand{\ee}{\end{equation}}
\def\tr{\mathop{\rm tr}\nolimits}

\def\dif{\mathop{\rm d\!}\nolimits}

\def\r{{\rm r}}

\begin{document}

\title[Labeling spherically symmetric spacetimes]
{Labeling spherically symmetric spacetimes with the Ricci tensor}

\author{Joan Josep Ferrando$^{1,2}$\
and Juan Antonio S\'aez$^3$}

\address{$^1$\ Departament d'Astronomia i Astrof\'{\i}sica, Universitat
de Val\`encia, E-46100 Burjassot, Val\`encia, Spain}

\address{$^2$\ Observatori Astron\`omic, Universitat
de Val\`encia, E-46980 Paterna, Val\`encia, Spain}

\address{$^3$\ Departament de Matem\`atiques per a l'Economia i l'Empresa,
Universitat de Val\`encia, E-46071 Val\`encia, Spain}

\ead{joan.ferrando@uv.es; juan.a.saez@uv.es}

\begin{abstract}
We complete the intrinsic characterization of spherically symmetric solutions partially accomplished in a previous paper [Class.Quant.Grav. (2010) {\bf 27} 205024]. In this approach we consider every compatible algebraic type of the Ricci tensor, and we analyze specifically the conformally flat case for perfect fluid and Einstein-Maxwell solutions. As a direct application we obtain the {\em ideal} labeling (exclusively involving explicit concomitants of the metric tensor) of the Schwarzschild interior metric and the Vaidya solution. The Stephani universes and some significative subfamilies are also characterized. 
\end{abstract}

%

\

\section{Introduction}
\label{sec-intro}

Despite the essential role played by spherically symmetric spacetimes in developing the General Relativity theory, and the extensive literature on spherically symmetric solutions of Einstein equations, an algorithmic characterization of these spacetimes has not been accomplished.  
The intrinsic characterization of spherically symmetric spacetimes is an old problem established and partially solved by Takeno in 1952 \cite{takeno} (see also \cite{takeno-2}). 

In \cite{fs-SSST} we have studied the Takeno problem and provided its solution for a broad family of metrics: those that are not conformally flat. Our study is based on the understanding of the algebraic structure of a type D Weyl tensor \cite{petrov} \cite{bel} \cite{fms}. Nevertheless, the Ricci tensor of a spherically symmetric spacetime always admits a space-like two-eigenplane and this property enables us to look for an alternative intrinsic characterization, which will be the only valid one in the conformally flat case. Thus, a deep knowledge of the algebraic structure of the Ricci tensor \cite{churchill} \cite{Plebanski} \cite{bcm} plays a central role.

In his paper Takeno \cite{takeno} referred to characterization in "ideal form" when a spacetime is labeled by equations exclusively involving explicit concomitants of the metric tensor. 
As Takeno claimed, the ideal characterization of spacetimes provides
an algorithmic way to test if a metric tensor, given in an arbitrary
coordinate system, is a specific solution of Einstein equations.
This can be useful to check whether a new solution is in fact already
known. The possibility of making the ideal characterization for any given metric is a consequence of the seminal results by Cartan \cite{cartan}. The analysis of the metric equivalence problem in General Relativity (see for example \cite{karlhede}) is based on the Cartan approach, which consists in working in an orthonormal (or a null) frame, fixed by the underlying geometry of the Riemann tensor. But in labeling a spacetime we can also use tensorial (non necessarily scalar) concomitants of the curvature tensor. This is the  approach that we apply in solving the Takeno problem, an approach that has been useful in labeling, among others, the Schwarzschild \cite{fsS}, Reissner-Nordstr\"om \cite{fsD} and Kerr \cite{fsKerr} black holes, or the Lema\^itre-Tolman-Bondi \cite{fs-SSST} and Bertotti-Robinson \cite{fswarped} solutions (see references therein for more examples).  These studies are of interest in obtaining a fully algorithmic characterization of the initial data which correspond to a given solution \cite{garcia-parrado-vk} \cite{garcia-parrado} \cite{garcia-parrado-2016}.

The paper is organized as follows. In section \ref{sec-ssst} we summarize some known results \cite{fs-SSST} on the intrinsic characterization of the spherical symmetry required in this paper. Section \ref{ideal-labeling} is devoted to labeling spherically symmetric spacetimes by using only the Ricci tensor when the spacetime is not an Einstein space. We consider the seven compatible Segr\'e types grouped in two classes: when the Ricci tensor admits a strict space-like two-eigenplane or when it admits a strict three-eigenplane. In section \ref{sec-weyl=0} we analyze in detail the conformally flat case by separately studying the Segr\'e types that include three physically relevant cases: perfect fluid, non null electromagnetic field and null electromagnetic field or pure radiation. In section \ref{sec-examples} we use our above results to label specific known families of solutions: the Stephani universes, the Schwarzschild interior solution and the Vaidya spacetime.

We also present three appendices. In the first one we summarize some of Takeno's results on invariant tensors in spherically symmetric spacetimes. The second one is devoted to presenting a new version to the Rainich theory for perfect fluid solutions, and we give specific statements when local thermal equilibrium or energy conditions are required; we use these results in sections \ref{sec-weyl=0} and \ref{sec-examples}.  In the third appendix we obtain the canonical form of some conformally flat metrics admitting a three-dimensional isometry group on two-dimensional space-like orbits.

In this paper we work on an oriented spacetime with a metric tensor
$g$ of signature $\{-,+,+,+\}$. The Weyl tensor $W$, the Ricci tensor $R$
and the scalar curvature $r$ are defined as given in \cite{kramer}. For the metric product of two vectors, we write $(x,y) =
g(x,y)$, and we put $x^2 = g(x,x)$. The symbols $\nabla$, $\nabla
\cdot$ and $\Delta$ denote, respectively, the covariant derivative,
the divergence operator and the Laplacian operator. For a 2-tensor $S$, $S^2= S \cdot S$ denotes the 2-tensor $(S^2)_{\alpha \beta} =  S_{\alpha}^{\ \lambda} S_{\lambda \beta}$, and $2 (S,S) = S_{\alpha \beta}S^{\alpha \beta}$.


\section{Characterizing spherically symmetric spacetimes}
\label{sec-ssst}

From the usual Takeno's \cite{takeno-2} canonical form for the metric line element of a spherically symmetric spacetime, we obtain \cite{fs-SSST} :
\begin{lemma} \label{lemma-can-ssst}
A spacetime is spherically symmetric if, and only if, the metric
tensor can be written as:
\begin{equation} \label{ss-can-2}
g = v + e^{2 \lambda} \hat{h} \, , \qquad \hat{h}(\dif \lambda)=0 \, ,
\end{equation}
where $v$ is an arbitrary {\rm 2}-dimensional Lorentzian metric, and
$\hat{h}$ is a {\rm 2}-dimensional Riemannian metric with a Gauss
curvature equal to one.
\end{lemma}

This lemma implies that the spacetime is a 2+2 s-warped product with {\it warped factor} $\lambda$ \cite{fswarped}. There are two principal planes, the time-like V with projector $v$ and the space-like H with projector $h =e^{2 \lambda} \hat{h}$. The pair $(v,h)$ defines a  2+2 almost-product structure with structure tensor $\Pi = v-h$. The {\it canonical} time-like
unitary 2-form $U$ is the volume element of the plane V, $U^2=v$. If we define $\hat{v} = e^{-2 \lambda}v$, we have $g = e^{2 \lambda}(\hat{v} + \hat{h})$, that is, the metric is conformal to a product one with a conformal factor which fulfills the second condition in (\ref{ss-can-2}).


\subsection{Invariant characterization of the spherically symmetric
spacetimes} \label{subsec-invariant-ssst}

From the invariant properties of the 2+2 warped products studied in \cite{fswarped}, we have obtained in \cite{fs-SSST} two different invariant characterizations of the spherically symmetric spacetimes, one based on conditions for the canonical unitary two-form $U$, and the other one based on conditions for the space-like projector $h$. The first one has been used in \cite{fs-SSST} in looking for a intrinsic labeling based on the Weyl tensor. The second one, which we use in this work, can be stated as follows \cite{fs-SSST}.
\begin{proposition} \label{prop-ssst-inv}
A spacetime is spherically symmetric if, and only if, there exists a space-like {\rm 2}-projector $h$ ($h^2 = h\, ,
\ \tr h = 2 \, , \ h(x,x) \geq 0 \, ,$ where $x$ in an arbitrary
time-like vector)  such that
\begin{equation} \label{s-warped-invariant-h}
2 \, \nabla h = h \stackrel{{\stackrel{23}{\sim}}}{\otimes} \xi  \,  ,
\qquad \dif \xi = 0  \,  , \qquad \xi \equiv  \nabla \! \cdot \! h  \, .
\end{equation}
\begin{equation} \label{kappa-ssst-invar}
\kappa \equiv \kappa(h) \, > \, 0 \, , \qquad h (\dif \kappa) = 0 \, ,
\end{equation}
where $\kappa(h)$ is the Gauss curvature of $h$, and where $(h \stackrel{{\stackrel{23}{\sim}}}{\otimes} \xi)_{\alpha \beta \gamma} = h_{\alpha \beta} \xi_{\gamma} + h_{\alpha \gamma} \xi_{\beta}$ .\\
Moreover $h$ is the projector on the {\rm 2}-spheres. The warping factor $\lambda$ can be obtained
as $\lambda = - \frac12 \ln \kappa$, and $\hat{h} = \kappa \, h$ is
the metric of the {\rm 2}-sphere.
\end{proposition}


\subsection{Ricci and Weyl tensors}
\label{subsec-ricweyl-warped}

Taking into account how the Weyl and Ricci
tensors change by a conformal transformation, we have \cite{fs-SSST}:
\begin{proposition} \label{prop-Ricci-weyl-1}
Let $g =e^{2 \lambda } \, ( \hat{v} + \hat{h} )=v + h$ be a spherically symmetric metric and let $X$ and  $\kappa$ be the Gauss curvatures of $\hat{v} $ and $h$, respectively. Then,
\begin{enumerate}
\item The Weyl tensor is type D (or O) with double eigenvalue $\varphi
\equiv - \frac{1}{6}\kappa (X+1)$. In the type D case, the
canonical {\rm 2}-form $U$, $U^2 = v$, is the principal {\rm 2}-form
of the Weyl tensor which takes the expression:
\begin{equation} \label{Weyl-conf-pro}
W = \rho(3 S + G) \, ,  \quad S \equiv U \otimes U -
*U\otimes *U  \, , 
\end{equation}
where $*$ is the Hodge dual operator and $G$ is the induced metric on the 2-forms space, $G_{\alpha \beta \lambda \mu} = g_{\alpha  \lambda} g_ {\beta \mu} - g_{\alpha  \mu} g_ {\beta \lambda}$.
\item
The space-like principal plane is an eigenplane of the Ricci tensor
$R$, that is, $R \cdot h = \mu \, h$ where $\mu$ is the associated
eigenvalue. The Ricci tensor takes the expression:
\begin{equation} \label{ricci-warped}
R = X \hat{v} + \hat{h} +  \nabla \xi - \frac12  \xi \otimes \xi + \frac12  ( \nabla \cdot \xi - \xi^2 ) \, {g} \, ,
\end{equation}
where $\xi =\nabla \cdot h$. Moreover, the scalar curvature $\kappa$ of $h$ admits the expressions:
\begin{equation} \label{gauss}
\kappa =  e^{- 2 \lambda} = \mu - \frac12  \nabla \cdot \xi =  - 2 \, \varphi- \frac{1}{6}\, \r + \mu + \frac14 \, \xi^2 \, .  
\end{equation}
\end{enumerate}
\end{proposition}
%


\subsection{Intrinsic characterization of the spherically symmetric spacetimes} 
\label{subsec-intrinsic-ssst}

Proposition \ref{prop-ssst-inv} characterizes the spherically
symmetric spacetimes in terms of a projector $h$ whatever their Riemann algebraic type. On the other hand, proposition \ref{prop-Ricci-weyl-1} states that $h$ must be the projector on a space-like eigenplane of the Ricci tensor, and $\kappa$ can be obtained in terms of this projector and its associated eigenvalue $\mu$. Thus, we have the following.
\begin{theorem} \label{theo-ssst-inv}
A spacetime is spherically symmetric if, and only if, the projector $h$ on a space-like eigenplane of the Ricci tensor $R$ satisfies
\begin{equation} \label{invariant-h}
2 \, \nabla h = h \stackrel{{\stackrel{23}{\sim}}}{\otimes} \xi \,  ,
\qquad \dif \xi = 0  \,  , \qquad \xi \equiv  \nabla \! \cdot \! h  \, .
\end{equation}
\begin{equation} \label{kappa-ssst-invar}
\kappa  > \, 0 \, , \qquad h (\dif \kappa) = 0 \, , \qquad \kappa \equiv  
- 2 \, \varphi- \frac{1}{6}\, \r + \mu + \frac14 \, \xi^2 \, ,
\end{equation}
where $\varphi$ is the double Weyl eigenvalue, $r = \tr R$, and $\mu$ is the eigenvalue associated with $h$.
Then, the metric $g$ with spherical symmetry admits the canonical form:
\begin{equation} \label{ss-can-3}
g = v + \frac{1}{\kappa} \, \hat{h} \, ,
\end{equation}
where $v=g-h$, and $\hat{h} = \kappa \, h$ is
the metric of the {\rm 2}-sphere.
\end{theorem}
{\it Remark 1} The approach presented in this section and therein is valid for any spacetime admitting a three-dimensional isometry group $G_3$ on 2-dimensional space-like orbits. In theorem above, the first condition in (\ref{kappa-ssst-invar}) can be changed to $\kappa <0$ (resp. $\kappa =0$) in order to characterize the hyperbolic (resp. plane) symmetry.


\section{Ideal labeling of the spherical symmetry by using the Ricci tensor}
\label{ideal-labeling}

After theorem \ref{theo-ssst-inv}, in order to obtain an ideal labeling (from the Takeno's point of view) of the spherically symmetric spacetimes, we must obtain, when possible, the projector $h$ in terms of the Ricci tensor. Of course, in the case of vacuum and Einstein spaces we must necessarily use the Weyl tensor to characterize spherical symmetry. Otherwise, the expression of $h$ in terms of the Ricci tensor depends on the Ricci algebraic type. In some (regular) types, $h$ can be obtained algebraically. Nevertheless, when algebraic degeneracies arise, we must use differential Ricci concomitants. Now we consider the different Ricci types which are compatible with spherical symmetry, that is, those admiting a space-like two-eigenplane. From now on, $N$ denotes the traceless part of the Ricci tensor, and $b,c,d$ the trace of the powers of $N$:
\begin{equation} \label{traces}
r \equiv \tr R , \quad b \equiv \tr N^2  , \quad c \equiv \tr N^3  , \quad d \equiv \tr N^4  , \quad   N = R - \frac14 r g  . 
\end{equation}
%


\subsection{Ricci tensor admitting a strict space-like two-eigenplane} 
\label{segre-2-plane}

This case corresponds with four possible Segr\'e algebraic types, $[z \tilde{z} (11)]$, $[1,1(11)]$, $[(1,1) (11)]$, $[2(11)]$, and can be distinguished from the case of a three-eigenplane by condition \cite{bcm} $7 b^2 \not= 12 d$. From the algebraic study of the symmetric tensors presented in \cite{bcm}, we obtain the following.

\begin{proposition} \label{prop-plane}
Let $R$ be a symmetric tensor, and consider its traceless part $N$ and the invariant algebraic scalars $r,b,c,d$ defined in {\em (\ref{traces})}, and let us define the algebraic concomitant:
\begin{equation} \label{P}
\hspace{-2.2cm}
P \equiv N^2 + 2 \nu N + (3 \nu^2 - \frac{b}{2}) g   ,  \qquad \nu \equiv \cases{ \frac{c(7b^2 - 12 d)}{3[b(b^2-4d)+4c^2]}, &   if \ \ 
$b^2\not=4d$ \cr \pm \frac{\sqrt{b}}{2}, & if \ \ $b^2=4d$ \cr }
\end{equation}
The symmetric tensor $R$ admits a (strict) two-eigenplane if, and only if, the concomitant $P$ satisfies
\begin{equation}
P^2 = \frac12 (\tr P) P \not = 0 \, .
\end{equation}
Then the projector on the two-plane and the associated double eigenvalue are given, respectively, by 
\begin{equation}
h = \frac{2}{\tr P} P \, , \qquad \qquad \mu = \nu + \frac14 r  \, .
\end{equation} 
Moreover the two-plane is space-like if, and only if, for any arbitrary time-like vector $x,$ 
\begin{equation} \label{spacelike}
h(x,x) \geq 0 \, .
\end{equation} 
\end{proposition}
The above proposition can also be inferred by showing that the imposed conditions are equivalent to $R$ taking the expression $R = S + \mu h$, with $S \cdot h = 0$.  

Now, from proposition \ref{prop-plane} and theorem \ref{theo-ssst-inv} we obtain:
\begin{theorem} \label{theo-ssst-ideal-two}
Let  $R$ be the Ricci tensor of a metric $g$, and let $r$, $b$, $c$, $d$, $N$, $\nu$ and $P$ the Ricci concomitants defined in {\em (\ref{traces})} and {\rm (\ref{P})}. When $7 b^2 \not= 12 d$, the necessary and sufficient conditions for $g$ to be spherically symmetric is that $\tr P \not=0$ and $h = \frac{2}{\tr P} P$ and $\mu = \nu + \frac14 r$ satisfy the algebraic conditions $h^2=h$ and {\rm (\ref{spacelike})}, and the differential ones {\rm (\ref{invariant-h})} and {\rm (\ref{kappa-ssst-invar})}.
\end{theorem}
%
%
{\it Remark 2} This theorem applies for the four Segr\'e types quoted above, and it admits a slightly different alternative statement, which avoids condition (\ref{spacelike}) in some cases. The sign of the invariant scalar $I \equiv 2b - \sqrt{|7b^2 - 12d|}$ allows us to discriminate some of these types \cite{bcm}. In type $[1,1(11)]$ ($I >0$), we necessarily need to impose condition (\ref{spacelike}) in order to distinguish it from the Segr\'e type $[(1,1)11]$, which is not compatible with spherical symmetry. In type $[z \tilde{z} (11)]$ ($I < 0$), it is not necessary to impose (\ref{spacelike}) since the other algebraic conditions in the theorem imply that a two-eigenplane exists, and now it is certainly space-like. When two double eigenvalues exist, we have $I=0$, which is equivalent to $b^2=4d$ under the established algebraic constraints. Now (\ref{P}) offers two possible amounts for the eigenvalue $\nu$. In theorem \ref{theo-ssst-ideal-two} one of the two amounts must accomplish the conditions in order to have spherical symmetry. Nevertheless, in type $[(1,1) (11)]$ ($N^2 = \frac{b}{4} g$), if (\ref{spacelike}) does not hold for one of the eigenvalues $\nu$, then it necessarily holds for the other eigenvalue $- \nu$. And in type $[2(11)]$, the space-like condition (\ref{spacelike}) holds for one of the eigenvalues when the other algebraic conditions hold for this eigenvalue.


\subsection{Ricci tensor admitting a strict three-eigenplane} 
\label{segre-3-plane}

This case corresponds with three possible Segr\'e algebraic types, $[(1,11)1]$, $[1,(111)]$, $[(211)]$, and it can be distinguished from the case of a strict two-eigenplane by condition $7 b^2 = 12 d$ \cite{bcm}. Now, from the results in \cite{bcm} we can obtain the following.

\begin{proposition} \label{prop-3-plane}
For a symmetric tensor $R$, we consider its traceless part $N$ and the invariant algebraic scalars $r,b,c,d$ defined in {\em (\ref{traces})}, and let us define the algebraic concomitant:
\begin{equation} \label{Q}
Q \equiv N +  \frac12 \sqrt[3]{\frac{c}{3}} \, g  \, .
\end{equation}
The symmetric tensor $R$ admits a (strict) three-eigenplane $\pi$ if, and only if, the concomitant $Q$ satisfies
\begin{equation} \label{3Q}
Q^2 =  (\tr Q) Q \, , \qquad Q\not=0 \, .
\end{equation}
Moreover $Q$ projects any vector $x$ on (i) a simple eigendirection orthogonal to $\pi$ when $ c \not=0$, (ii) the null eigendirection lying on $\pi$ when $c=0$. In both cases, this invariant vector and the eigenvalue associated with $\pi$ are given, respectively, by: 
\begin{equation} \label{x-mu} 
e = \frac{1}{\sqrt{Q(x,x)}} Q(x),  \qquad  \mu = - \frac12 \sqrt[3]{\frac{c}{3}} + \frac14 r  \, ,
\end{equation} 
where $x$ is any arbitrary vector such that $Q(x) \not=0$.
\end{proposition}
Again, this proposition can also be proved by considering the canonical form of the three considered Segr\'e types.

Proposition \ref{prop-3-plane} provides necessary conditions and gives the multiple eigenvalue $\mu$ and an invariant vector $e$. Then, in order to determine the projector $h$ on the space-like principal plane we can make use of differential concomitants of the Ricci tensor which can be obtained from $\mu$, $e$ and the scalar curvature $r$. From lemmas \ref{lemmaTak(invariants)} and \ref{lemmah(invariants)} in \ref{Ap-invariances}, and taking into account that $v=g-h= U^2$, we obtain:

\begin{proposition} \label{prop-3-plane-F}
In a spherically symmetric spacetime with a Ricci tensor admitting a three-eigenplane, let $r$, $\mu$ and $e$ be the algebraic Ricci invariants considered in proposition \ref{prop-3-plane} and let us consider the differential invariants:
\begin{equation} \label{F}
\hspace{-1cm} F_{(1)} \equiv \dif r \wedge e  , \qquad F_{(2)} \equiv \dif \mu \wedge e  , \qquad F_{(3)} \equiv \dif  e \, . 
\end{equation}
If one of these invariants, say $F_{(A)}$, does not vanish, the projector $h$ on the space-like principal plane can be obtained as:
\begin{equation} \label{h(F)}
h = g-v \, , \qquad v = -\frac{1}{(F_{(A)}, F_{(A)})} F_{(A)} \cdot F_{(A)} \, . 
\end{equation} 
\end{proposition}

Lemma \ref{lemmah(invariants)} in \ref{Ap-invariances} implies that the invariant vector $e$ is always hypersurface orthogonal, and when $\dif e =0$, one has $S = \alpha f \otimes f$, with $f$ a vector in the time-like principal plane and which is orthogonal to $e$: $h(f)=0$, $(e,f)=0$. Then,  if $e$ is a null vector $\ell$ (Segr\'e type [(211)]), we have $f=\ell$ and $\ell$ is tangent to a non-rotating shear-free geodesic null congruence. Consequently, 
\begin{equation} \label{nabla-ell}
\nabla \ell =  \rho \ell \otimes \ell + \frac12 \theta h \, , \qquad \theta = \nabla \cdot \ell  \, . 
\end{equation} 
When $\theta = 0$, the Ricci identities for $\ell$ and the Bianchi identities imply that $\ell$ is parallel to a covariantly constant null vector. Otherwise, we have:
\begin{proposition} \label{prop-3-null}
In a spherically symmetric spacetime with a Ricci tensor admitting a three-eigenplane, let $e$ be the Ricci eigenvector considered in proposition \ref{prop-3-plane}. If all of the invariants $F_{(A)}$ in {\em (\ref{F})} vanish and $e=\ell$ is a null vector such that $\theta \not=0$, the projector $h$ on the space-like principal plane can be obtained as:
\begin{equation} \label{ell-h(theta)}
h = \frac{4}{\theta^2} \nabla \ell \cdot \nabla \ell \, . 
\end{equation} 
\end{proposition}

On the contrary, if $e^2\not=0$, we can consider the unitary vector $u= |e^2|^{-\frac12} e$, $u^2 = \epsilon=\pm1$. Then $f^2 = -\epsilon |f^2|$, and we can obtain the projector $h$ when $e$ is not tangent to a shear-free congruence. Moreover, we can look for a second order differential concomitant of the Ricci tensor obtained from the expansion of $u$. Indeed, after a simple algebraic calculation we obtain:
\begin{proposition} \label{prop-3-sigma}
In a spherically symmetric spacetime with a Ricci tensor admitting a three-eigenplane, let $e$ be the Ricci eigenvector considered in proposition \ref{prop-3-plane}. If all the invariants $F_{(A)}$ in {\em (\ref{h(F)})} vanish and $e^2 \not= 0$, let $\gamma$, $\theta$ and $\sigma$ be the projector orthogonal to $e$, the expansion and the shear:
\begin{equation} \label{gamma-sigma}
\hspace{-10mm} \gamma = g - \epsilon u \otimes u , \quad \sigma = \nabla u - \frac13 \theta \gamma , \quad \theta = \nabla \cdot u  , \quad  u= \frac{e}{\sqrt{|e^2|}}  , \quad \epsilon = u^2  . 
\end{equation} 
Then, if $\sigma \not=0$, the projector $h$ on the space-like principal plane can be obtained as:
\begin{equation} \label{h(sigma)}
h = \frac23 \gamma  - \sqrt[3]{\frac{2}{9 \tr \sigma^3}} \sigma  \, . 
\end{equation} 
And, if $F \equiv \dif \theta  \wedge e \not=0$, the projector $h$ on the space-like principal plane can be obtained as:
\begin{equation} \label{h(dtheta)}
h = g-v \, , \qquad v = -\frac{1}{(F, F)} F \cdot F   \, . 
\end{equation}  
\end{proposition}

Segr\'e type [(211)] corresponds to the case $c=0$, $Q^2=0$ and $e^2 = 0$ in proposition \ref{prop-3-plane}. When the first derivatives of the Ricci tensor can not determine the projector $h$ by using propositions \ref{prop-3-plane-F} and \ref{prop-3-null}, then a covariantly constant null vector exists, and the Bianchi and Ricci identities lead to $r=0$ and $W=0$. Then, the spacetime is a plane wave (which is not spherically symmetric). 

Segr\'e types [1,(111)] and [(1,11)1] correspond to $c\not=0$, $\tr Q\not=0$ and $e^2 \not= 0$ in proposition \ref{prop-3-plane}. When the first derivatives of the Ricci tensor and the gradient of the expansion can not determine the projector $h$ by using propositions \ref{prop-3-plane-F} and \ref{prop-3-sigma}, then a vorticity-free and shear-free geodesic non null congruence with homogeneous expansion exists. Then, it is known (and can easily be deduced from the Bianchi and Ricci identities) that the spacetime is hypersurface-homogeneous. If $e^2>0$ (type [1,(111)]) the spacetime is a Lemaître-Friedmann-Robertson-Walker universe (spherically symmetric). If $e^2 <0$ (type [(1,11)1]) the spacetime is not spherically symmetric. 

From these considerations, propositions \ref{prop-3-plane}, \ref{prop-3-plane-F}, \ref{prop-3-null}, \ref{prop-3-sigma},  and theorem \ref{theo-ssst-inv}, we obtain:
\begin{theorem} \label{theo-ssst-ideal-three}
Let  $R$ be the Ricci tensor of a metric $g$, and let $r$, $b$, $c$, $d$, $N$, $Q$, $e$ and $\mu$ the Ricci concomitants defined in {\em (\ref{traces}), (\ref{Q})} and {\em (\ref{x-mu})}. When $7 b^2 = 12 d$, a necessary condition for $g$ to be spherically symmetric is that $Q$ satisfies the algebraic condition {\em (\ref{3Q})}. Under these restrictions we have:

If one of the invariants $F_{(A)}$ given in {\em (\ref{F})} does not vanish, the necessary and sufficient conditions for $g$ to be spherically symmetric is that $(F_{(A)}, F_{(A)})<0$ and $\mu$ and $h$ given by {\em (\ref{h(F)})} satisfy the differential conditions {\em (\ref{invariant-h})} and {\em (\ref{kappa-ssst-invar})}, and $h^2=h$, $\tr h =2$.

If $F_{(A)}=0$ for all $A$, and $e^2=0$, the necessary and sufficient conditions for $g$ to be spherically symmetric is that $\theta \not=0$ and that $\mu$ and $h$ given by {\em (\ref{ell-h(theta)})} satisfy the differential conditions {\em (\ref{invariant-h})} and {\em (\ref{kappa-ssst-invar})}, and $h^2=h$, $\tr h =2$, $h(x,x) \geq 0$, for any time-like vector $x$.

If $F_{(A)}=0$ for all $A$, $e^2\not=0$, and $F=\dif \theta \wedge e \not =0$, the necessary and sufficient conditions for $g$ to be spherically symmetric is that $(F, F)<0$ and that $\mu$ and $h$ given by {\em (\ref{h(dtheta)})} satisfy the differential conditions {\em (\ref{invariant-h})} and {\em (\ref{kappa-ssst-invar})}.

If $F_{(A)}=0$ for all $A$, $\sigma \not=0$, and $e^2<0$, the necessary and sufficient conditions for $g$ to be spherically symmetric is that $\mu$ and $h$  given by  {\em (\ref{h(sigma)})} satisfy the differential conditions {\em (\ref{invariant-h})} and {\em (\ref{kappa-ssst-invar})}, and $h^2=h$. When $F_{(A)}= F=\sigma=0$, and $e^2<0$,  the spacetime is a Lema\^itre-Friedmann-Robertson-Walker universe (spherically symmetric). 

If $F_{(A)}=0$ for all $A$, $F=0$, and $e^2>0$, the necessary and sufficient conditions for $g$ to be spherically symmetric is that $\sigma \not=0$, and that $\mu$ and $h$ given by  {\em (\ref{h(sigma)})} satisfy {\em (\ref{spacelike})} and the differential conditions {\em (\ref{invariant-h})} and {\em (\ref{kappa-ssst-invar})}, and $h^2=h$, $h(x,x) \geq 0$, for any time-like vector $x$.  
\end{theorem}
%
%


\section{Labeling conformally flat spherically symmetric spacetimes}
\label{sec-weyl=0}

Theorems \ref{theo-ssst-ideal-two} and \ref{theo-ssst-ideal-three} characterize all the spherically symmetric metrics by only using the Ricci tensor. When the Weyl tensor does not vanish we can acquire other simpler characterizations involving the Weyl tensor or involving both, the Ricci and Weyl tensors \cite{fs-SSST}. But in the conformally flat case we must necessarily draw on the Ricci tensor. Nevertheless, some conditions in the above theorems become identities when the Weyl tensor vanishes. Now we analyze this fact for the Segr\'e types of the Ricci tensor that include the physically more relevant cases: perfect fluid, null and non null electromagnetic field and pure radiation.

\subsection{Ricci tensor of Segr\'e type {\em[1,(111)]}} 
\label{W0-fluper}

In this case, the Ricci tensor $R$ has an associated perfect energy tensor. Now we work, as usual, with the hydrodynamic variables $(u,\rho,p)$:
\begin{equation} \label{T-fluper}
R - \frac{1}{2} r \, g = T \equiv  (\rho + p) \, u \otimes u + p g \, .
\end{equation}

The Bianchi identities take the expression $\nabla \! \cdot \! W = B$ where $B$ is the Cotton tensor, $B_{\alpha \beta \gamma} \equiv \nabla_{[\alpha} D_{\beta]\gamma}$, $2 D \equiv R - \frac{r}{6} g$. In the conformally flat case, $W=0$, the Cotton tensor vanishes, a condition which is equivalent to the following constraints on the hydrodynamic variables:
\begin{equation} \label{W=0}
\nabla u = - u \otimes a + \frac13 \theta \gamma   , \quad \dif \rho = (\rho + p)\theta \, u  , \quad \dif p + \dot{p} u + (\rho+p) a =0 ,
\end{equation}
where a dot means directional derivative with respect to the unit velocity $u$. Then, the Ricci identities for $u$ lead to:
\begin{equation} \label{Ricci-iden}
\hspace{-24mm} 
\dif \theta \! \wedge \! u =0  , \quad \nabla a = \frac{\alpha}{3} \gamma - a \! \otimes \! a -u \otimes \dot{a} + \frac{\theta}{3} a \otimes u    , \quad \nabla \! \cdot \! a =  \alpha \equiv \dot{\theta} + \frac{1}{3} \theta^2 + \frac{\rho \! + \! 3p}{2}  \, .
\end{equation}
When $a=0$, the spacetime is a LFRW universe. Otherwise, when $a \not=0$, the projector on the time-like plane defined by the spherical symmetry is:
\begin{equation} \label{v}
v= - u \otimes u + \Omega \, a \otimes a , \qquad \Omega \equiv \frac{1}{a^2} = -\frac{1}{(u,\dot{a})} . 
\end{equation}
From (\ref{Ricci-iden}), we have:
\begin{equation} \label{dOmega}
\dif  \Omega = 2   \Omega^2 (a, \dot{a}) \, u + 2 \Omega (1- \frac13  \alpha \Omega)\, a \,  .
\end{equation}
Then, from this expression and (\ref{Ricci-iden}) and (\ref{v}), we obtain:
\begin{equation} \label{nabla-v}
\nabla h = - \nabla v=  \Omega \, u \otimes a \stackrel{\sim}{\otimes} h(\dot{a}) + \frac12 h \stackrel{{\stackrel{23}{\sim}}}{\otimes} \xi, \qquad \xi \equiv \frac23 [  \theta u - \Omega \alpha a ]  .
\end{equation}

From (\ref{Ricci-iden}) we obtain $\dif \dot{\theta} \wedge u \wedge a = 0$. Consequently, $h(\dif \dot{\theta})=0$ and then $h( \dif \alpha)=0$. Moreover, we have $9 \xi^2 = 4(\Omega \alpha^2 - \theta^2)$. Thus $h(\dif \xi^2) =0$. This condition, (\ref{W=0}) and (\ref{Ricci-iden}) imply that $h(\dif \kappa)=0$,  where $\kappa$ is given in (\ref{kappa-ssst-invar}).  

On the other hand, from (\ref{invariant-h}) and (\ref{nabla-v}) we establish that $h(\dot{a})=0$ is a necessary condition for spherical symmetry. Now we show that this condition implies $\dif \xi =0$.

From (\ref{Ricci-iden}), (\ref{dOmega}) and $h(\dif \alpha)=0$, we obtain that $\dif \xi=0$ is equivalent to the scalar condition:
\begin{equation} \label{dz=0}
 \Omega (\dot{\alpha} + \frac13 \theta \alpha) = \theta + \alpha \Omega^2(a, \dot{a})  .
\end{equation}
If we differentiate $h(\dot{a})=0$ and we make use of (\ref{nabla-v}), we have:
\begin{equation} \label{nabla-a-punt}
h^{\alpha \beta}\nabla_{\alpha} \dot{a}_{\beta} = -\dot{a}^{\beta}\nabla_{\alpha} h_{\beta}^\alpha = -(\dot{a},\xi)  =  \frac{2}{3 \Omega} [\theta + \alpha \Omega^2 (a, \dot{a})]   .
\end{equation}
Moreover, if we make use of (\ref{Ricci-iden}), the Ricci identities for $a$ lead to $h^{\alpha \beta}\nabla_{\alpha} \dot{a}_{\beta} = 2(\dot{\alpha} + \frac13 \theta \alpha)$. Thus (\ref{dz=0}) holds and $\dif \xi =0$.

Then, from theorem \ref{theo-ssst-inv}, and taking into account the relation between the Ricci and $T$ scalar invariants, $r = \rho - 3p$, $2 \mu = \rho-p$, we obtain the following intrinsic characterization:

\begin{proposition} \label{prop-intrinsec-fluper}
A conformally flat perfect fluid solution is spherically symmetric if, and only if, the hydrodynamic variables $(u,\rho, p)$ satisfy either $a=0$, or $a\not=0$ and
\begin{equation} \label{intrinsec-fluper}
a \wedge \dif a = 0 \, ,\qquad \kappa >0 \, .
\end{equation}
where
\begin{equation} \label{intrinsec-fluper-def}
a \equiv \nabla_u u  , \qquad \kappa \equiv \frac16 (3 \rho -p) + \frac19 \! \left[\frac{(\nabla \! \cdot \! a)^2}{a^2} \!- \! \theta^2\right]  .
\end{equation}
\end{proposition}
In order to obtain an ideal characterization of the spherically symmetric conformally flat perfect fluid solutions, we must obtain the intrinsic conditions in the proposition above in terms of explicit concomitants of the Ricci tensor.

From theorem \ref{theo-fluper} in \ref{Ap-Rainich-fluper} and proposition \ref{prop-intrinsec-fluper}, we establish that a conformally flat spacetime is a spherically symmetric perfect fluid solution  if, and only if, the Ricci tensor satisfies conditions (\ref{fluper-conditions-A}), (\ref{fluper-definitions-A}) and the hydrodynamic variables $(u, \rho,p)$ given in (\ref{fluper-hydro}) are submitted to conditions (\ref{intrinsec-fluper}), (\ref{intrinsec-fluper-def}). The expression of $u$ in (\ref{fluper-hydro}) involves an arbitrary vector $x$. In order to prevent this $x$ appearing in equations (\ref{intrinsec-fluper}), we can give the expressions (\ref{intrinsec-fluper-def}) in terms of Ricci concomitants by using the projector $\Gamma \equiv \frac{1}{s}Q = u \otimes u$. Note that $a = q + \Gamma(q)$, and $\theta^2 = \Gamma(q,q)$, where $q = \nabla\! \cdot \! \Gamma$. Then, we finally obtain:
\begin{theorem} \label{theo-cf-fluper}
A spacetime is a spherically symmetric conformally flat perfect fluid solution if, and only if, the Weyl tensor vanishes, $W=0$, and the Ricci tensor $R$ satisfies conditions
\begin{equation} \label{fluper-conditions}
Q^2 + s Q = 0  , \qquad s Q(x,x) > 0   ,
\end{equation}
where $x$ is any time-like vector, and 
\begin{equation} \label{fluper-definitions}
Q \equiv N - \frac14 s g , \quad N \equiv R - \frac14 r g   , \quad s \equiv - 2 \sqrt[3]{\frac{\tr N^3}{3}} , \quad    r\equiv \tr R ,
\end{equation}
and, either $a=0$, or $a\not=0$ and {\rm (\ref{intrinsec-fluper})}, where:
\begin{equation} \label{intrinsec-fluper-def-Gamma}
\hspace{-20mm} a \equiv q + \Gamma(q)  , \quad \kappa \equiv \frac16 (2s+r) + \frac19 \! \left[\frac{(\nabla \! \cdot \! a)^2}{a^2} \!- \! \Gamma(q,q)\right] , 
 \quad q \equiv  \nabla\! \cdot \! \Gamma , \quad  \Gamma \equiv \frac{1}{s}Q \, .
\end{equation}
\end{theorem}
{\it Remark 3} It is known \cite{st} \cite{kramer} that the conformally flat perfect fluid solutions are, either the Stephani universes (when $\theta \not=0$) or the generalized Schwarzschild interiors (when $\theta=0$). In both families there are spherically symmetric solutions and, consequently, they are contained in the set of metrics characterized in the theorem above. Nevertheless, in order to obtain this theorem, we have preferred to follow a self-contained reasoning without using any previously known result. In section \ref{sec-examples} we come back to the conformally flat perfect fluid solutions, and we offer the ideal characterization of some known specific solutions. 

The result in theorem (\ref{theo-cf-fluper}) can be extended to the metrics admitting a $G_3$ on space-like two-dimensional orbits by removing the second condition in (\ref{intrinsec-fluper}). Thus, we can state:

\begin{proposition} \label{prop-G3-fluper}
A spacetime is a conformally flat perfect fluid solution admitting a group $G_3$ on space-like two-dimensional orbits if, and only if, the Weyl tensor vanishes, $W=0$, and the Ricci tensor $R$ satisfies conditions {\em (\ref{fluper-conditions})} and $a \wedge \dif a = 0$, where $Q$, $s$ and $r$ are given in {\rm (\ref{fluper-definitions})}, $a$ is given in {\em (\ref{intrinsec-fluper-def-Gamma})}, and $x$ is any vector such that $Q(x) \not= 0$.
\end{proposition}
%


\subsection{Ricci tensor of Segr\'e type {\em[(1,1)(11)]}} 
\label{W0-cem}

Now we have a time-like eigen-two-plane with projector $v$ and a space-like eigen-two-plane with projector $h$. If $\Pi = v-h$ is the structure tensor, the Ricci tensor takes the form:
\begin{equation} \label{Ricci2+2}
R= \chi \Pi + \frac14 r g \, .
\end{equation}
When $W=0$, the Bianchi identities state that the Cotton tensor vanishes, a condition which is equivalent to:
\begin{eqnarray}
12 \,h(\dif \chi ) = h (\dif  r) \, , \qquad 12 \, v(\dif \chi ) = -v (\dif  r) \, ,   \label{bianchi-1}  \\ 
2 \nabla  h =   h \stackrel{\stackrel{23}{\sim}}{\otimes} \xi  +  v \stackrel{\stackrel{23}{\sim}}{\otimes}   \zeta \, , \quad \xi \equiv 2  v(\dif \ln \chi) \, , \quad  \zeta \equiv 2 h(\dif \ln \chi) \, .
\label{bianchi-2}
\end{eqnarray}
These expressions and the Ricci identities for $h$ lead to:
\begin{equation} \label{Ricci-iden-2+2}
\dif \xi^2 = 2 \nabla \xi \cdot \xi = (\nabla \cdot \xi + \frac32 \xi^2 ) \xi \,  .
\end{equation}
Now the scalar invariant $\kappa$ given in (\ref{kappa-ssst-invar}) is $12 \kappa = r - 12 \chi + 3 \xi^2$, and (\ref{bianchi-1}) and (\ref{Ricci-iden-2+2}) imply $h(\dif \,[r -12 \chi])=0$ and $h(\dif \xi^2)=0$. Consequently, 
we have $h(\dif \kappa)=0$. 

On the other hand, from (\ref{invariant-h}) and (\ref{bianchi-2}) we have that $\zeta=0$ (that is, $h(\dif \chi)=0$), is a necessary condition for spherical symmetry. But this conditions implies $v(\dif \chi) = \dif \chi$ so that $\dif \xi =0$. Then, from theorem \ref{theo-ssst-inv} we obtain the following intrinsic characterization:
\begin{proposition} \label{prop-intrinsec-2+2}
A conformally flat metric with a Ricci tensor of type {\em [(1,1)(11)]} is spherically symmetric if, and only if, the Ricci-invariants $(\Pi, r , \chi)$ satisfy $\Pi(\dif \chi) = \dif \chi$ and 
\begin{equation} \label{intrinsec-2+2}
\kappa \equiv \frac{1}{12} r - \chi + (\dif \ln \chi)^2 >0  \, .
\end{equation}
\end{proposition}
In order to obtain an ideal characterization of the spherically symmetric conformally solutions of type [(1,1)(11)] we must obtain the intrinsic conditions in the proposition above in terms of explicit concomitants of the Ricci tensor. 

The Ricci algebraic type can be guaranteed by imposing the algebraic Rainich conditions on the Ricci traceless part $N= \chi \Pi$. In order to obtain the sign of the scalar invariant $\chi$ we can use the sign of $N(x,x)$, $x$ being a time-like vector. Then, we finally obtain:
\begin{theorem} \label{theo-cf-2+2}
A spacetime is a spherically symmetric conformally flat solution with Ricci tensor of type {\em [(1,1)(11)]} if, and only if, the Weyl tensor vanishes, $W=0$, and the Ricci tensor $R$ satisfies conditions {\em (\ref{intrinsec-2+2})} and 
\begin{equation} \label{2+2-conditions}
N^2 = \chi^2 g \not=0  , \qquad N(\dif \chi) = \dif \chi   ,
\end{equation}
where
\begin{equation} \label{2+2-definitions}
N \equiv R - \frac14 r g , \quad    r\equiv \tr R    , \quad \chi \equiv \frac{\epsilon}{2} \sqrt{\tr N^2}     , \quad \epsilon \equiv -\frac{N(x,x)}{|N(x,x)|}  ,
\end{equation}
where $x$ is any time-like vector.
\end{theorem}
{\it Remark 4} Note that in our study above we have not imposed, neither $r$ is constant, nor the Maxwell equations. Under the conformally flat hypothesis, these two conditions are equivalent. Indeed, relations (\ref{bianchi-1}) imply that $\dif r =0$ if, and only if, $\dif \chi =0$, which is tantamount to the conservation condition for $N$. Moreover (\ref{bianchi-2}) implies that $\nabla v=0$ and then the metric is a product one \cite{fswarped}. Then, the differential Rainich conditions identically hold and $N$ is the energy tensor of non null Maxwell filed. In fact, the metric is the Bertotti-Robinson Einstein-Maxwell solution with cosmological constant $\Lambda = \frac14 r$. The ideal characterization of this solutions is known \cite{kramer} \cite{fswarped}.
\ \\[2mm]
{\it Remark 5} For the sake of completeness we have included in theorem \ref{theo-cf-2+2} also the case of $\dif r \not=0$, which evidently does not represent a Einstein-Maxwell solution, as commented above. The canonical form of these metrics has been obtained in \ref{Ap-forma-canonica}. It is given by expressions (\ref{G3S2}), (\ref{vhat}), and (\ref{H-2+2}), taking $\epsilon =1$.

On the other hand, the result in theorem \ref{theo-cf-2+2} can be extended to the metrics admitting a $G_3$ on space-like two-dimensional orbits by removing condition (\ref{intrinsec-2+2}). Thus, we can state:

\begin{proposition} \label{prop-G3-2+2}
A spacetime is a conformally flat solution with Ricci tensor of type {\em [(1,1)(11)]} and admitting a group $G_3$ on space-like two-dimensional orbits if, and only if, the Weyl tensor vanishes, $W=0$, and the Ricci tensor $R$ satisfies conditions {\em (\ref{2+2-conditions})}, where $N$ and $\chi$ are given in {\rm (\ref{2+2-definitions})}, and $x$ is any time-like vector.
\end{proposition}
%


\subsection{Ricci tensor of Segr\'e type {\em[(211)]}} 
\label{W0-radiacio}

In this case we have a null eigenvector $\ell$ and the Ricci tensor takes the form 
\begin{equation}\label{ems1}
R= \epsilon \ell \otimes \ell + \frac14 r g  , \qquad \epsilon^2 =1 .
\end{equation} 

If $\dif r = 0$, when $W=0$ the Bianchi identities imply that a vector $e$  exists such that
\begin{equation}
\dif \ell = \ell \wedge e , \qquad  (e,\ell)=0 .
\label{ems-0}
\end{equation}
Consequently, the invariant $\dif \ell $ is a null two-form. But, when a $G_3$ on two-dimensional space-like orbits exists, this result is inconsistent with lemma \ref{lemmaTak(invariants)} in \ref{Ap-invariances}. Then, necessarily $e=0$ and Bianchi identities implies that $\nabla \ell = z \ell \otimes \ell$. Moreover, Ricci identities for $\ell$ lead to $\dif z \wedge \ell =0$ and, consequently, a null vector collinear with $\ell$ exists which is covariantly constant. Thus, we can state:
\begin{proposition} \label{prop-211-0}
The conformally flat metrics admitting three-dimensional isometry group on two-dimensional space-like orbits, and with a Ricci tensor of type {\em [(211)]} and $\dif r=0$, are plane waves. Consequently, they are never spherically symmetric.
\end{proposition}

If $\dif r \neq 0$, when $W=0$ the Bianchi identities imply that a null vector $k$ with $(k, \ell)=-1$ and three scalars $z$, $\beta$, $\tau$,  exist such that
\begin{eqnarray}
\nabla \ell = z \ell  \otimes \ell - \beta k  \otimes \ell + 2 \beta
h, \qquad \dif r = 24 \epsilon \beta \ell \label{ems2}, \\
\dif \beta = \beta^2 k + \tau \ell , \qquad \dif k = \ell \wedge
\Big[ \frac{1}{\beta^2} \dif \tau - 3 \frac{\tau}{\beta} k \Big],
\label{ems3}
\end{eqnarray}
where $h = g + \ell \stackrel{\sim}{\otimes} k $ is the projector onto the orthogonal plane defined by $\dif \ell$. Using the
expression above for $\nabla \ell$, the Ricci identities for $\ell$ lead to:
\begin{eqnarray}
\hspace{-1.5cm}  2 \nabla h = h \stackrel{\stackrel{23}{\sim}}{\otimes} \xi + \ell
\otimes \ell \stackrel{\stackrel{23}{\sim}}{\otimes} \zeta ,\quad \xi \equiv
\frac{1}{\beta}\left(  \frac{r}{12}+ 2 \tau - 2 \beta  z
  \right) \ell + 4 \beta k, \quad \zeta \equiv \frac{2}{\beta} h(\dif z),
  \label{ems4} \\
\hspace{-1.0cm}   h(\dif z)= \frac{1}{\beta} h(\dif \tau) , \qquad \ell \wedge  \Big[ v(\dif z) +
  \frac{1}{\beta} v(\dif \tau) \Big] =  \Big[  \beta z + 2 \tau - \frac{r}{12}
  \Big]\ell \wedge k ,  \label{ems5}
\end{eqnarray}
where $v = - \ell \stackrel{\sim}{\otimes} k = g-h$. Then, from (\ref{invariant-h}) and (\ref{ems4}), a necessary condition for spherical symmetry is that $h(\dif z)=0$, which is equivalent to $h(\dif \tau)=0$ if we take into account the first equation in (\ref{ems5}). Moreover, under this condition we obtain $\dif \xi   =0$ as a consequence of (\ref{ems2}), (\ref{ems3}) and (\ref{ems4}) .

On the other hand, putting $\varphi=0$, $\mu = \frac14 r$ and the value of $\xi^2$ obtained from (\ref{ems4}) in the expression (\ref{kappa-ssst-invar}) for the curvature of $h$, we obtain $\kappa = \frac{1}{16} r + \frac{1}{2} (\beta z - \tau) $, and then $h(\dif \kappa)=0$ as a consequence of (\ref{ems2}), (\ref{ems3}) and (\ref{ems5}). Then, we obtain $h(\dif z)=0$ or, equivalently, $2 \nabla h = h \stackrel{\stackrel{23}{\sim}}{\otimes} \xi $, with $\xi \equiv \nabla \cdot h$, is a sufficient condition for the spacetime to admit a three-dimensional isometry group a G$_3$ on space-like two-dimensional orbits. Thus, from theorem \ref{theo-ssst-inv} we obtain the following intrinsic characterization:
\begin{proposition} \label{prop-intrinsec-211}
A conformally flat metric with a Ricci tensor of type {\em [(211)]} is spherically symmetric if, and only if, the Ricci-invariants $(r, h , \xi)$ satisfy 
\begin{eqnarray} \label{intrinsec-211}
\dif r \not=0, \qquad  \qquad  2 \nabla h = h \stackrel{\stackrel{23}{\sim}}{\otimes} \xi,    \\
\label{intrinsec-211-kappa}
\kappa \equiv \frac{1}{12}r +  \frac14 \xi^2 >0  \, .
\end{eqnarray}
\end{proposition}
In order to obtain an ideal characterization of the spherically symmetric conformally solutions of type [(211)] we must obtain the intrinsic conditions in the proposition above in terms of explicit concomitants of the Ricci tensor. 

The Ricci algebraic type can be guaranteed by imposing the algebraic Rainich conditions on the Ricci traceless part $N= \epsilon \ell \otimes \ell$. Furthermore, from (\ref{ems2}) and (\ref{ems3}) we can determine the Hessian and the Laplacian of the scalar $r$ as:
\begin{equation} \label{Hessiana}
\nabla \! \dif r = 24 \epsilon (\tau + \beta z) \ell \otimes \ell + 48 \epsilon  \beta^2 h, \qquad \Delta r = 96 \epsilon \beta^2 .
\end{equation}
These expressions allow us to obtain the projector $h$ in terms of the Ricci tensor. Then, we finally have:
\begin{theorem} \label{theo-cf-211}
A spacetime is a spherically symmetric conformally flat solution with Ricci tensor of type {\em [(211)]} if, and only if, the Weyl tensor vanishes, $W=0$, and the Ricci tensor $R$ satisfies $N^2 =0$ and conditions {\em (\ref{intrinsec-211})} and {\em (\ref{intrinsec-211-kappa})}, where
\begin{equation} \label{211-definitions}
N \equiv R - \frac14 r g , \quad    r\equiv \tr R    , \quad h \equiv \frac{4}{(\Delta r)^2} \nabla \! \dif r \cdot \nabla \! \dif r    , \quad \xi \equiv \nabla \cdot h  .
\end{equation}
\end{theorem}
{\it Remark 6} Under the conformally flat condition, we have that $\dif r=0$ if, and only if, $\nabla \cdot N =0$. Consequently, proposition \ref{prop-211-0} and theorem \ref{theo-cf-211} impliy that the radiation field $N= \epsilon \ell \otimes \ell$ is never the energy tensor of a null electromagnetic field. The canonical form of these metrics has been obtained in \ref{Ap-forma-canonica}. It is given by expressions (\ref{G3S2}), (\ref{vhat}), and (\ref{H-ll-not0}), taking $\epsilon =1$.
\ \\[2mm]
{\it Remark 7}  From (\ref{ems2}) we establish that the two-form $F \equiv \dif \ell \neq 0$ is orthogonal to the isometry group orbits. Moreover, if $x$ is any time-like vector, $N(x)$ is collinear with $\ell$. This way, we can determine the projector $h$ on the orbits in a similar way as in (\ref{h(F)}):
\begin{equation} \label{211-h-ell}
h = g-v  , \quad v = -\frac{1}{(\dif \ell, \dif \ell)} \dif \ell \cdot \dif \ell  , \quad \ell \equiv \frac{\epsilon N(x)}{\sqrt{|N(x,x)|}}   .
\end{equation}
This expression for $h$ is an alternative to that given in (\ref{211-definitions}).
\ \\[2mm]
{\it Remark 8} It follows easily from (\ref{ems3}) that the second condition in (\ref{intrinsec-211}) is equivalent to the integrability, $\dif k \wedge k =0$, of the null vector $k$ defined by $\dif \ell = \beta \ell \wedge k$. This property is similar to that assuring the existence of a G$_3$ in the perfect fluid case, $\dif a \wedge a =0$, where the acceleration $a$ is defined by $\dif u = u \wedge a$.

On the other hand, the result in theorem \ref{theo-cf-211} can be extended to the metrics admitting a $G_3$ on space-like two-dimensional orbits by removing condition (\ref{intrinsec-211}). Now we must add the case $\dif r=0$ that leads to plane waves (proposition \ref{prop-211-0}), which admit plane symmetry.  Thus, we can state:
\begin{proposition} \label{prop-G3-211}
A spacetime is a conformally flat solution with Ricci tensor of type {\em [(211)]} and admitting a group $G_3$ on space-like two-dimensional orbits if, and only if, the Weyl tensor vanishes, $W=0$, and the Ricci tensor $R$ satisfies $N^2 =0$ and, either it is a plane wave, or conditions {\em (\ref{intrinsec-211})} hold, where $N$, $r$, $h$ and $\xi$ are given in {\em(\ref{211-definitions})}.
\end{proposition}
%


\section{Labeling specific known solutions}
\label{sec-examples}

The characterizations of the spherically symmetric metrics presented in previous sections have been built without reference to any already known specific solution. Now we focus on labeling some renowned solutions: the Stephani universes, the Schwarzschild interior and the Vaydia metric.

\subsection{Ideal characterization of the Stephani universes}
\label{subsec-stephani}

As commented in remark 3, the Stephani universes are the expanding conformally flat perfect fluid solutions \cite{st} \cite{kramer}. Under these hypotheses, (\ref{W=0}) implies that $\theta =0$ is equivalent to $\dif \rho =0$. Then, from theorem \ref{theo-fluper} in \ref{Ap-Rainich-fluper} we have:
\begin{theorem} \label{theo-stephani}
The Stephani universes are characterized by the following conditions:
\begin{equation} \label{stephani-conditions}
W=0, \qquad Q^2 + s Q = 0  , \qquad s Q(x,x) > 0   , \qquad 3 \dif s +  \dif r \not=0 ,
\end{equation}
where $W$ and $R$ are, respectively, the Weyl and Ricci tensors, $Q$, $s$ and $r$ are given in {\rm (\ref{fluper-definitions})} and $x$ is any time-like vector. 
\end{theorem}
Moreover, the Stephani universes become the LFRW ones when $a=0$ or, equivalently, when the energy tensor is a barotropic perfect fluid, $\dif \rho \wedge \dif p=0$. Thus, taking into account (\ref{fluper-hydro}), we have:
\begin{corollary} \label{cor-LFRW}
The Lema\^itre-Friedmann-Robertson-Walker universes are characterized by conditions {\rm (\ref{stephani-conditions})} and $\dif r \wedge \dif s = 0$.
\end{corollary}

Bona and Coll \cite{bc} showed that the Stephani universes admitting a thermodynamic scheme are those admitting a $G_3$ on two-dimensional orbits. Then, from proposition \ref{prop-G3-fluper} we can state:
\begin{corollary} \label{cor-stephani-termo}
We have the following equivalent conditions:\\[1mm]
(i) The spacetime is a Stephani universe representing a fluid evolving in local thermal equilibrium. \\[1mm]
(ii) The spacetime is a Stephani universe admitting a $G_3$ on two-dimensional orbits. \\[1mm]
(iii) The metric tensor $g$ fulfills {\em (\ref{stephani-conditions})} and $a \wedge \dif a =0$, where $W$ and $R$ are, respectively, the Weyl and Ricci tensors, $Q$, $s$ and $r$ are given in {\rm (\ref{fluper-definitions})}, $a$ is given in  {\em (\ref{intrinsec-fluper-def-Gamma})} and $x$ is any time-like vector. 
\end{corollary}
{\it Remark 9}: Corollary above states the already known \cite{bc} equivalence between a physical property (point (i)) and a geometric one (point (ii)), and offer their ideal characterization (point (iiii)). Note that we have added a purely kinematic condition, $a \wedge \dif a =0$, to theorem \ref{theo-stephani}. We could replace this additional second-order kinematic condition for the hydrodinamic one (\ref{thermo-R-gamma}), which is a first-order condition in the velocity $u$ and second-order one in the scalar invariants.
\ \\[2mm]
{\it Remark 10}: Each of the statements in corollary \ref{cor-stephani-termo} includes the well defined LFRW limit: (i) the thermodynamic scheme degenerates to a barotropic evolution, $\dif r \wedge \dif s = 0$; (ii) the isommetry group expands to a $G_6$ on three-dimensional orbits; (iii) the kinematic constraint becomes $a=0$.


\subsection{Ideal characterization of the interior Schwarzschild solution}
\label{subsec-interior}

The non expanding conformally flat perfect fluid solutions are the so-called {\em generalized Schwarzschild interiors} \cite{st} \cite{kramer}. They can be characterized by a similar statement to that of theorem \ref{theo-stephani} by changing the last condition in (\ref{stephani-conditions}) for $3 \dif s +  \dif r =0$. The Schwarzschild interior is the static spherically symmetric subclasse. The static constraint implies that $\dif a=0$, which entails the first condition in (\ref{intrinsec-fluper}), $a \wedge \dif a=0$. On the other hand, $\theta=0$ implies the second condition in (\ref{intrinsec-fluper}), $\kappa > 0$, provided that the energy conditions hold. Also, note that now $a=q$ where $q$ is given in (\ref{intrinsec-fluper-def-Gamma}). Thus, we can state:
\begin{theorem} \label{theo-interior}
A spacetime is the Schwarzschild interior solution if, and only if, the metric tensor fulfills the following conditions:
\begin{eqnarray} \label{interior-1}
W=0, \qquad Q^2 + s Q = 0  , \qquad  Q(x,x) > 0   , \\
\label{interior-2}   s > 0   , \quad  r+s > 0 ,  \quad   3 \dif s + \dif r = 0 , \quad \dif \left[\nabla \cdot (s^{-1}Q)\right] =0 ,
\end{eqnarray}
$W$ and $R$ being, respectively, the Weyl and Ricci tensors, and where $x$ is any time-like vector, and $Q$, $r$ and $s$ are given in {\rm (\ref{fluper-definitions})} in terms of the Ricci tensor $R$.
\end{theorem}
%


\subsection{Ideal characterization of the Vaidya solution}

Vaidya spacetime is the only spherically symmetric solution with Ricci tensor of type [(211)], constant scalar curvature, and submitted to the energy conditions. Thus, the Ricci tensor is given by:
\begin{equation} \label{va1}
R= \epsilon \ell \otimes \ell + \frac14 r g , \qquad \dif r =0, \qquad \epsilon >0 .
\end{equation}
Different expressions for this solution can be found in literature (see, for example \cite{kramer}). It is a type D solution, and we could use the results in \cite{fs-SSST} in order to intrinsically characterize the metric by using the Weyl tensor. Here we draw on the Ricci tensor $R$ and the sole algebraic scalar defined by the Weyl tensor, its double eigenvalue $\varphi$.

In the Vaidya spacetime we have that the two-form $F \equiv \dif \varphi \wedge \ell \neq 0$ is orthogonal to the isometry group orbits. Moreover $(\dif \varphi, \ell) \not=0$, and the energy conditions imply $N(\dif \varphi, \dif \varphi) >0$, where $N\equiv R - \frac{1}{4} r g = 
\ell \otimes \ell$. Then $\ell = N(\dif \varphi)/\sqrt{N(\dif \varphi, \dif \varphi)}$. This way, we can determine the projector $h$ on the orbits in a similar way to how we proceeded in (\ref{h(F)}), and now we can obtain it in terms of $N$ and $\dif \varphi$.

On the other hand, theorem \ref{theo-ssst-inv} states that (\ref{invariant-h}) are necessary conditions for spherical symmetry. From these conditions and the Ricci identities for $h$ we can deduce that $h(\dif \xi^2)=0$. Consequently, the scalar curvature $\kappa$ given by (\ref{kappa-ssst-invar}) satisfies $h(\dif \kappa) =0$ provided that $h$ fulfills (\ref{invariant-h}). Then, aplying theorem \ref{theo-ssst-inv} and considering the Rainich conditions for a Ricci tensor of type (\ref{va1}), we obtain the following characterization of the Vaidya solution.
\begin{theorem} \label{theo-Vaidya}
A spacetime is the Vaidya solution if, and only if, the Weyl tensor $W$ and the Ricci tensor $R$ fulfill the following conditions:
\begin{eqnarray} \label{va2}
A\not=0, \qquad N^2 = 0  , \qquad  N(\dif \varphi, \dif \varphi) > 0 , \qquad \dif r=0, \\
2 \nabla h = h \stackrel{\stackrel{23}{\sim}}{\otimes} \xi,  \qquad \dif \xi =0,   \qquad \kappa > 0,  \label{va3}
\end{eqnarray}
where 
\begin{eqnarray} \label{va4}
N\equiv R - \frac{1}{4} r g, \quad r \equiv \tr R,  \quad \varphi \equiv  \frac{2C}{A} , \quad A\equiv  \rm{tr} W^2, \quad C \equiv \rm{tr} W^3, \\
\label{va5}
h \equiv g-v, \quad v \equiv \frac{1}{N(\dif \varphi,\dif \varphi)}  \Big[ N(\dif \varphi) \stackrel{\sim}{\otimes} \dif \varphi - (\dif \varphi)^2  N \Big], \quad \xi \equiv \nabla \cdot h ,  \\
\kappa \equiv - 2 \varphi + \frac{1}{12} r + \frac{1}{4} \xi^2 . \label{va6}
\end{eqnarray}
\end{theorem}

\ack This work has been supported by the Spanish ``Ministerio de
Econom\'{\i}a y Competitividad", MICINN-FEDER project FIS2015-64552-P.

\appendix

\section{Invariant tensors in spherically symmetric spacetimes}
\label{Ap-invariances}

It is not difficult to extend a Takeno's result \cite{takeno-2} on spherically symmetric spacetimes:
\begin{lemma} \label{lemmaTak(invariants)}
Let $U$ and $h$ be the canonical two-form and the projector on the two-dimensional space-like orbits of a three-dimensional isometry group. If a vector $x$, a symmetric two-tensor $B$ and a two-form $F$ are invariant by the isometry group, then
\begin{equation}
\begin{array}{l}
h(x) = 0  , \qquad    B = S + \lambda h  ,  \qquad    F = \alpha U + \beta *\!U  ; \\[2mm]
S \cdot h = 0  , \qquad   h (\dif \lambda) = h(\dif \alpha) =h(\dif \beta) = 0  \, .
\end{array}
\end{equation} 
\end{lemma}
Then, one can easily obtain:
\begin{lemma} \label{lemmah(invariants)}
If $\rho$ is an invariant scalar and $x$ is an invariant vector, then:
\begin{equation}
\begin{array}{l}
h(\dif \rho) = 0 \, , \qquad \nabla x = \alpha U + S + \beta h \, ;   \\[2mm]
h(\dif \alpha) =h(\dif \beta) = 0 , \qquad  S \cdot h = 0 ,  \qquad  ^tS = S \, . 
\end{array} 
\end{equation} 
\end{lemma}
%


\section{Rainich conditions for a perfect fluid solution}
\label{Ap-Rainich-fluper}

The necessary and sufficient conditions for a metric $g$ to be a non null Einstein-Maxwell solution were studied by Rainich \cite{rainich}. A similar approach for the perfect fluid implies providing the conditions for a Ricci tensor to take the form (\ref{T-fluper}). Explicit conditions for a symmetric tensor to be of perfect fluid type are known \cite{bcm} \cite{Torre}. Here we present a simpler equivalent version of these results which can be deduced from proposition \ref{prop-3-plane} by considering the case $e^2 <0$. Note that now $Q(x)\not=0$ for any time-like vector $x$. Then, we have::
\begin{theorem} \label{theo-fluper}
A spacetime is a perfect fluid solution if, and only if, the Ricci tensor $R$ satisfies:
\begin{equation} \label{fluper-conditions-A}
Q^2 + s Q = 0  , \qquad s Q(x,x) > 0   ,
\end{equation}
where $x$ is any time-like vector, and 
\begin{equation} \label{fluper-definitions-A}
Q \equiv N - \frac14 s g , \quad N \equiv R - \frac14 r g   , \quad s \equiv - 2 \sqrt[3]{\frac{\tr N^3}{3}} , \quad    r\equiv \tr R .
\end{equation}
The total energy $\rho$, the pressure $p$ and the unit velocity $u$ of the fluid are given by:
\begin{equation} \label{fluper-hydro}
 \rho = \frac14 (3 s + r)   , \qquad p = \frac14 (s - r)  , \qquad  u = \frac{Q(x)}{\sqrt{s Q(x,x)}}   .
\end{equation}
\end{theorem}
If our goal is to obtain solutions representing a physically realistic fluid we must impose complementary constraints, like the energy conditions or the existence of a local thermal equilibrium scheme. 

The perfect fluid energy tensors that represent the energetic evolution of a perfect fluid in local thermal equilibrium can be characterized by the hydrodynamic constraint \cite{Coll-Ferrando-termo} \cite{fluperLTE} : $(\dot{\rho} \dif \dot{p}- \dot{p} \dif \dot{\rho})  \wedge \dif \rho \wedge \dif p =0$.
Then, from theorem \ref{theo-fluper} we arrive to the Rainich-like theory for the thermodynamic perfect fluid:
\begin{theorem} \label{theo-fluper-thermo}
A spacetime is a thermodynamic perfect fluid solution if, and only if, the Ricci tensor $R$ satisfies {\rm (\ref{fluper-conditions-A}), (\ref{fluper-definitions-A})} and 
\begin{equation} \label{thermo-R}
(\dot{s} \dif \dot{r}- \dot{r} \dif \dot{s})  \wedge \dif r \wedge \dif s =0 .
\end{equation}
where a dot denotes directional derivative with respect to the $u$ given in {\em (\ref{fluper-hydro})}.
\end{theorem}
Equation (\ref{thermo-R}) needs the expression (\ref{fluper-hydro}) for $u$, which involves an arbitrary vector $x$. Nevertheless, taking into account that $Q=s u\otimes u$ we can replace (\ref{thermo-R}) by the condition:
\begin{equation} \label{thermo-R-gamma}
Q( \dif r) = 0    \qquad  {\rm or}  \qquad   Q( \dif r) \not= 0 , \quad    \dif \left[\frac{Q(\dif s, \dif r)}{Q(\dif r, \dif r)}\right]  \wedge \dif r \wedge \dif s =0 .
\end{equation}

The Pleba\'nsky \cite{Plebanski} energy conditions for a perfect fluid energy tensor state $-\rho< p \leq \rho$. Then, from theorem \ref{theo-fluper} we obtain:
\begin{theorem} \label{theo-fluper-energy}
A spacetime is a perfect fluid solution subjected to the energy conditions if, and only if, the Ricci tensor $R$ satisfies:
\begin{equation} \label{fluper-conditions-A2}
Q^2 + s Q = 0  , \qquad  Q(x,x) > 0   ,  \qquad  s > 0   , \qquad  r+s > 0   ,
\end{equation}
where $x$ is any time-like vector, and $Q$, $r$ and $s$ are given in  {\rm (\ref{fluper-definitions-A})}.
\end{theorem}

Finally, from the two last theorems we obtain an improved version of a result in \cite{Coll-Ferrando-termo}:
\begin{theorem} \label{theo-fluper-energy-thermo}
A spacetime is a thermodynamic perfect fluid solution subjected to the energy conditions if, and only if, the Ricci tensor $R$ satisfies {\rm (\ref{thermo-R-gamma})} and {\rm (\ref{fluper-conditions-A2})}, where $x$ is any time-like vector, and $Q$, $r$ and $s$ are given in  {\rm (\ref{fluper-definitions-A})}.
\end{theorem}
%


\section{Canonical form of some conformally flat metrics}
\label{Ap-forma-canonica}

In section \ref{sec-weyl=0} we have offered the ideal labeling of the conformally flat spherically symmetric spacetimes for the Segr\'e types [1,(111)], [(1,1) (11)] and [(211)]. The explicit coordinate expression of some of these solutions are known. The metric canonical form of the Stephani universes and the generalized Schwarzschild interior can be found in literature \cite{kramer} (Segr\'e type [1,(111)]). Here we give the canonical form for the other two Segr\'e types. 

As explained in section \ref{sec-ssst}, a metric admitting a $G_3$ on $S_2$ takes the expression:
\begin{equation} \label{G3S2}
g = \frac{1}{H^2}  \, [ \hat{v} + \hat{h}] \, , \qquad  \hat{h}(\dif H) = 0 \, ,
\end{equation}
where $\hat{h}$ is a two-dimensional Riemannian metric of constant curvature $\epsilon = 0, 1, -1$. If $g$ is a conformally flat metric, then (see proposition \ref{prop-Ricci-weyl-1}) the 2-dimensional lorentzian metric $\hat{v}$ is of constant curvature $-\epsilon$. Consequently, they admit the known canonical form:
\begin{equation} \label{vhat}
\hspace{-23mm} \hat{v}= \frac{1}{ \left( 1 - \frac{\epsilon}{2} uv \right)^2 } \dif u
\stackrel{\sim}{\otimes} \dif v  , \quad  \hat{h} = \frac{1}{\left( 1 +  \frac{\epsilon}{4} r^2 \right)^2 } ( \dif x  \otimes \dif x  + \dif y \otimes \dif y) , \ \  r^2 \equiv x^2 + y^2  .
\end{equation}
Then, (\ref{G3S2}) states that $H=H(u,v)$, and the Ricci tensor takes the expression:
\begin{equation} \label{ricciG3S2} \label{ri2} 
R=   2 \left[- \epsilon \hat{v}  +  \frac{1}{H} \nabla \dif H
\right] + \left[ \epsilon + \frac{\Delta H}{H} - 3 \frac{(\dif H)^2}{H^2} 
 \right] (\hat{v} + \hat{h}) ,
 \end{equation}
where now $\nabla$ and $\Delta$ denote the covariant derivative and the Laplacian operator of the metric $\hat{v}$, which can be evaluated taking into account that:
\begin{equation}\label{nabla-dv}
\nabla \dif u = \frac{- \epsilon v}{1 - \frac{\epsilon}{2} u v } \dif u
\otimes \dif u , \qquad \nabla \dif v = \frac{-\epsilon u}{1 -
\frac{\epsilon}{2} u v} \, \dif v \otimes \dif v .
\end{equation}
%


\subsection{Case {\em [(1,1)(11)]}}
\label{Ap-forma-canonica-2+2}

Now $R= \chi \Pi + \frac14 r g$. Then from (\ref{ricciG3S2}) and (\ref{nabla-dv}) we obtain:
\begin{equation} \label{H-2+2}
H=\frac{k_1 u v + k_2 u + k_3 v + k_4}{(1-\frac{\epsilon}{2}uv)} \, , 
\end{equation}
\begin{equation} \label{Lambda-2+2}
 \chi = (k_1 - \frac{\epsilon}{2} k_4) H  , \quad r = 12 \left[ (\frac{\epsilon}{2} k_4 - k_1 ) H + 4\,
\frac{1- \epsilon u v }{2 - \epsilon u v} (k_1 k_4 - k_2 k_3 ) \right] ,
\end{equation}
where $k_i$, $i=1,2,3,4$, are arbitrary constants such that $2 k_1 \not= \epsilon k_4$.


\subsection{Case {\em [(211)]}}
\label{Ap-forma-canonica-ll}

Now $R= \pm \ell \otimes \ell + \frac14 r g$. The null eigen-vector $\ell$ necessarily lies on the time-like plane and, consequently, it is proportional to the gradient of one of the coordinates $\{u,v\}$, say $u$. Then, $R= \lambda \dif u  \otimes \dif u+ \frac14 r g$, and from of (\ref{ricciG3S2}) and (\ref{nabla-dv}) we obtain:
\begin{itemize}
\item[-]
When $\epsilon =0$ 
\begin{equation} \label{H-ll-0}
H= k \, v + q (u) \, , \qquad    \lambda = \frac{q''(u)}{H}, \qquad r = - 24 k q'(u)   \, ,  
\end{equation}
where $k$ is an arbitrary constant and $q(u)$ an arbitrary function. 
\item[-]
When $\epsilon \not=0$ 
\begin{equation} \label{H-ll-not0}
H= \frac12 \left[ u q'(u) + \frac{2 + \epsilon u v}{2 -
\epsilon u v} q(u) \right] \, ,   
\end{equation}
\begin{equation} \label{H-Lambda-not0}
\hspace{-20mm} \lambda = \frac{1}{2 \, H} [ u q''' (u) + 3 q''(u) ] , \quad r= 3 \epsilon \Big[ (q(u) - u q'(u) )^2 - 2 u^2
q(u) q'' (u) \Big]   ,  
\end{equation}
where $q(u)$ is an arbitrary function. 
\end{itemize}


\end{document}